# CHARGE TO MAGNETIC FLUX RATIOS

**Gustavo González-Martín**

Departamento de Física, Universidad Simón Bolívar,

Apartado 89000, Caracas 1080-A, Venezuela.

Web page URL  http:\\prof.usb.ve\ggonzalm\

It is shown that the geometric assumption that the carriers in the fractional quantum Hall effect have integral multiples of charge quanta *e* and magnetic flux *h/2e*, as proposed in a previous publication dealing with a unified theory, is compatible with the experimental data.

## Magnetic Flux Quanta and the FQHE.

It is possible that magnetic flux quanta be associated to the electron orbits corresponding to definite total angular momentum levels in an electron gas in a constant intense magnetic field. This association may be helpful in analyzing the quantum Hall effect. The theory of the IQHE and the FQHE has evolved by the explicit construction of quantum state wave functions that describe many features of this phenomena [1,2,3]. Nevertheless it may still be possible to explain certain facts from general principles, as the extraordinary accuracy of the results suggests. This is the purpose of this paper. Of course, detailed microscopic analysis and wave functions are still necessary for a complete description. It would be like using the theory of representations of the Lorentz group to characterize the electron states in a central potential instead of using actual wave function equations. In this manner we expect to display the general principle involved. In fact, it was the generalization of the Lorentz group (automorphisms of Minkowsky space) to the group of automorphisms of the geometric algebra of Minkowsky space and the corresponding physical interpretation [4], that indicated that any particle (or quasi particle) is a representation of the structure group of the the unified theory and, therefore, must carry not only quanta of angular momentum, but also quanta of charge *e* and quanta of magnetic flux, *h/2e* , (*one or more*), both intrinsic and orbital. These requirements limits the value of flux to charge to certain ratios.

Usually the motion in a constant magnetic field is discussed in Cartesian coordinates in terms of states with definite energy and a definite linear momentum component. The resultant Landau energy levels are degenerate in terms of the momentum component. Additionally, for the electron, the Landau energy levels are doubly degenerate except the lowest level. In order to use definite angular momentum states, the problem is expressed in cylindrical coordinates and the energy of the levels is [5],

$$= \left(n+m+\tfrac{1}{2}+s\right)\frac{eB\hbar}{Mc} \; , \tag{1}$$

where *B* is the magnetic field along the positive *z* symmetry axis, *m* is the absolute value of the orbital angular momentum quantum number, also along the positive *z* axis, *n* is a non negative quantum number associated to the radial wave function and *s* is the spin. This expression has the peculiarity that even for zero *m* there may be energy quanta associated to the radial direction.

The equations of motion of charged particles in a constant magnetic field, in quantum or classical mechanics, do not determine the momentum or the center of rotation of the particle. These variables are the result of a previous process that prepares the state of the particle. We can idealize this process as a collision between the particle and the field. Since the magnetic field does not do work on the particle, the energy of the particle is conserved. The angular momentum of the particle with respect to the eventual center of motion is also conserved. The values of angular momentum and energy inside the field equal their values outside. This implies there is no kinetic energy associated to the radial wave function, as equation (1) allows. Therefore, among the possible values of energy we must exclude all values of *n* except the value zero. This means that the energy only depends on angular momentum, as in the classical theory.

The degeneracy of energy levels, apart from that due to the momentum component along the field, is the one due to the spin direction, as in the case of Cartesian coordinates. Only moving electrons make a contribution to the Hall effect, hence we disregard zero orbital angular momentum states. Each degenerate level has two electrons, one with spin down and orbital momentum *m+1* and the other with spin up and orbital angular momentum *m* . Each typical degenerate energy level may be considered to have a total charge *q* equal to *2e,* total spin *0* and a total integer orbital angular momentum *2m+1*. Using half integer units,

$$L_z = 2(2m+1)\left(\hbar/2\right) \tag{2}$$

In the classical treatment of motion in a magnetic field, it is known that the classical kinetic momentum of a particle has a circulation around the closed curve corresponding to a given orbit, that is twice the circulation of the canonical momentum which equals the negative of the external magnetic flux enclosed by the loop [6]. That is, for the given number of quanta of orbital angular momentum in a typical level, we may associate a number of orbital external magnetic flux quanta. In accordance with the principle indicated at the beginning, we assign a flux $\Phi_L$ to a degen-erate level,

$$\Phi_L = -2(2m+1)\left(h/2e\right) \qquad (3)$$

When we have an electron gas instead of a single electron, the electrons occupy a number of the available states in the energy levels depending on the Fermi level. The orbiting electrons are effective circular currents that induce a flux antiparallel to the external flux. The electron gas behaves as a diamagnetic body, where the applied field is reduced by the induced field to a net field because of the electron motion. Associated to the macroscopic magnetization $M$, magnetic induction $B$ and magnetic field $H$ we introduce, respectively, a magnetization flux $\Phi_M$, a net $B$ flux $\Phi_B$ and a bare $H$ flux $\Phi_H$ per level, which are related by

$$\Phi_B = \Phi_H + \Phi_M \; . \qquad (4)$$

The first energy level of moving electrons is non degenerate, corresponding to one electron with spin down and one quantum of orbital momentum. In order for the electron to orbit there has to be a net flux inside the orbit. If flux is quantized, the minimum possible flux is one quantum of orbital flux for the electron. In addition, another quantum is required by the intrinsic spin of the electron. Therefore without further equations, the number of net quanta in this level is 2 and its net magnetic flux is,

$$\Phi_B = (2)\frac{h}{2e} \; . \qquad (5)$$

The magnetization flux considering its induced part $\Phi_L$, 2 quanta for the orbital motion of the single electron, and its intrinsic part $\Phi_S$, one quantum, is,

$$\Phi_M = \Phi_L + \Phi_S = -2\left(\frac{h}{2e}\right) + \frac{h}{2e} = -\frac{h}{2e} \; , \qquad (6)$$

which gives, for this non degenerate level, the relation,

$$|\Phi_B| = 2|\Phi_M| \; , \qquad (7)$$

indicating an equivalent magnetic permeability of 2/3.

This reasoning is not applicable to the calculation of $\Phi_B$ for the degenerate levels in the electron gas, rather, we somehow have to find a relation with $\Phi_H$. To determine this exactly we would need to solve the (quantum) equation of motion for the electrons simultaneously with the (quantum) electromagnetic equation determining the field produced by the circulating electrons. Instead of detailed equations we recognize that, since flux is quantized, the orbital flux *must change by discrete quanta* as the higher energy levels become active. As the angular momenta of the levels increase, a proportional increase in any two of the variables in equation (4), would mean that the magnetic permeability does not vary from the value 2/3 determined by the non degenerate level. We choose as condition for the flux $\Phi_B$ that this ratio may deviate from the proportionality indicated by the last relation, by an integral number of flux quanta according to,

$$|\Phi_B| = 2|\Phi_M| + \Delta\Phi \; , \qquad (8)$$

where $\Delta\Phi$ is an indeterminate quantum flux of the electron pair. The flux $\Phi_M$ has $\Phi_L$ as upper bound,

$$|\Phi_M| \leq 2(2m+1)\left(h/2e\right) \; , \qquad (9)$$

and may write then,

$$|\Phi_B| \leq 2\left[2(2m+1)h/2e\right] + 2\delta h/e \; , \qquad (10)$$

where $\delta$ is an integer indicating a jump in the flux associated to each electron in the pair.

Whatever flux $\Phi_B$ results, it must be a function of only the flux quanta per electron pair. A degenerate energy level, consisting of the combination or coupling of two electrons in states of opposite spins, with orbital flux quanta $2\mu$ characterizing one electron and $2\mu+2$, characterizing the other electron of the pair, has $2(2\mu+1)$ quanta, in terms of a non negative integer $\mu$. Hence the only independent variable determining $\Phi_B$ is $2(2\mu+1)$. In order for $\Phi_B$ to be only a function of this variable, the integer $\delta$ must be even so that it adds to $m$, giving the possible values,

$$\Phi_B = 4(2\mu+1)\left(h/2e\right) \quad \mu \le m + \delta/2 \quad , \tag{11}$$

The number of possible quanta linked to this *superfluxed* level, indicated by $f$, is

$$f = 4(2\mu+1), \tag{12}$$

where $\mu$ has undetermined bound. The number of electric charge quanta indicated by $q$, is

$$q = 2 \; . \tag{13}$$

The general expression for the Hall conductivity, in terms of the two dimensional density of carriers $N/A$ and the number of quanta of the carriers $q$, is

$$\sigma = \frac{qeN}{AB} \; . \tag{14}$$

Each single electronic state is degenerate with a finite multiplicity. This gives the population of each typical energy half level corresponding to a single electron, indicated by $N_0$. The total electronic population when the levels are exactly full is,

$$qN = N_0 + 2\nu N_0 = (2\nu+1)N_o \; , \tag{15}$$

where $\nu$ doubly degenerate Landau energy levels are full and where the first $N_0$ corresponds to the non degenerate first Landau level. It is clear that if the degeneracy in the energy levels is lifted, by the mechanism described in the references, half levels may be filled separately and observed experimentally. The number $\nu$ is an integer if the highest full level is a complete level, a half integer if the highest full level is a half level and zero if the only full level is the non degenerate first level.

Since there is the same number of electrons $N_0$ in each sublevel we may associate one electron from each sublevel to a definite center of rotation, as a magnetic vortex. We have then, as model, a system of electrons rotating around a magnetic center, like a flat magnetic atom (quasi particle?). The validity of this vortex model relies in the possibility that the levels remain some how chained together, otherwise, the levels would move independently of each other. A physical chaining mechanism is clear: If we have two current loops linked by a common flux, and one loop moves reducing the flux through the other, Lenz's law would produce a reaction that opposes the motion of the first loop.

The flux linked to the system, whatever quantum theory wave functions characterize the states of the electronic matter, should have quanta because the system is a representation that must carry definite quanta of charge, angular momentum and magnetic flux. In other words, the magnetic vortex carries quanta of magnetic flux. The flux linking a particular vortex is equal to the net flux $\Phi_B$ linked by the orbit of the electron pair corresponding to the highest level. The value of $f$, the number of net quanta linked to a vortex, is expressed by equation (12) depending on the value of $\mu$ for the highest level in the system.

We now make the assumption that the condition for filling the Landau levels should be taken in the sense given in the previous article [4], by counting the *net* flux quanta linked to the orbital motion of all $N_0$ vortices. The filling condition is determined by conservation of flux (continuity of flux lines). The applied external flux must equal the total *net* flux linked *only* to the highest level. This determines a partial filling of the levels, in the canonical sense, where filling determines a flux of only $h/e$ per electron. This filling condition implies, for a degenerate level with $N_0$ pairs,

$$\frac{\Phi}{N_0} = f\left(\frac{h}{2e}\right). \tag{16}$$

Substituting, we obtain for the conductivity, for values of $q$ an $f$ given by equations (12, 13)

$$\sigma = \frac{e(2\nu+1)N_0}{\Phi} = \frac{2\nu+1}{2(2\mu+1)}\left(\frac{e^2}{h}\right) \quad \text{for } \nu \geq \tfrac{1}{2}. \tag{17}$$

This expression is not valid if the highest level is the non degenerate first level. For this case, where $\nu$ equal to zero, there are separate values for $q$ and $f$,

$$\frac{\Phi}{N_0} = f_0\left(\frac{h}{2e}\right), \tag{18}$$

and, we get, using equation (5),

$$\sigma = \frac{eN_0}{\Phi} = \frac{1}{f_0}\left(\frac{2e^2}{h}\right) = \frac{e^2}{h} \quad \text{for } \nu = 0. \tag{19}$$

For half integer $\nu$ we may define another half integer $n$ which indicates the number of full levels of definite angular momentum number $m$ (not energy levels), by

$$n = \nu + \frac{1}{2}. \tag{20}$$

If we replace $\nu$ by $n$ we get for the conductivity, the equivalent expression,

$$\sigma = \frac{n}{(2\mu+1)}\left(\frac{e^2}{h}\right) \quad \text{for } n \geq 1. \tag{21}$$

In this scheme, a highest full level is characterized by the integer $\mu$, characterizing the number of flux quanta per electron and either one of the half integers, $\nu$, number of active degenerate energy levels, or $n$, number of active orbital angular momentum levels. As the magnetic intensity is increased, the levels, their population and flux quanta are rearranged, to obtain full levels. Details for this process depends on the microscopic laws. Nevertheless, the quantum nature of magnetic flux requires a fractional *exact* value of conductivity whenever the filling condition is met independent of details. As the magnetic field increases the resultant fractions, for small integers $n$, $\mu$, are

$$\begin{aligned} &\cdots 3,\ 7/3,\ 2,\ 5/3,\ 7/5,\ 4/3,\ 6/5,\ 1,\ 6/7,\ 4/5,\ 7/9,\ 5/7, \\ &2/3,\ 7/11,\ 3/5,\ 4/7,\ 5/9,\ 6/11,\ 7/13,\ 6/13,\ 5/11,\ 4/9, \\ &3/7,\ 2/5, \cdots \end{aligned} \tag{22}$$

These fractions match the results of the odd fractional quantum Hall effect [7,8]. The same expression, for integer $\nu$, indicates plateaus at fractions with an additional divisor of 2. For example, if $\mu$ is zero, 3/2 for $\nu=1$, 5/2 for $\nu=2$, (but not ½ since $\nu>0$). This appears to be compatible with presently accepted values [9,10,11].

A partially filled level or superfluxed level occurs at a value of magnetic field which is fractionally larger than the corresponding value for a normal level with equal population because the carriers have fractionally extra flux.

The values of the conductivity are degenerate in the sense that one value corresponds to more than one set $\mu, \nu$. The magnetic field for any two sets with a given conductivity ratio is the same because the smaller number of carriers for one set is precisely compensated by the larger number of flux per carrier. Nevertheless, the two sets differ electromagnetically because of the extra flux per electron for one of the sets, and we may expect small energy differences between them, lifting this degeneracy. Of course, the proof of this difference would require detailed analysis using wave functions corresponding to a Hamiltonian that includes appropriate terms. If all electrons are in Landau levels, the Fermi level would jump directly from one Landau level to the next and the conductivity curve would be a set of singular points. Localized states due to lattice imperfections allow a Fermi level between Landau levels, as discussed in the references. If this is the case, their small separation would make the two sets coalesce into a plateau. Since the value of the conductivity, at two sets $\mu, \nu$ with the same fraction, is exactly the same, the value of the conductivity should be insensitive to a small variation of the magnetic field (or energy) indicating a finite width plateau at the value given by equation (17) with very great precision. In particular for the ratio 1 there are many low numbered sets that coalesce producing a very wide plateau. This accounts for the plateaus seen at the so called fractional and integral fillings.

This results may be understood in a different manner. Equation (16) is a consequence of the magnetic flux quanta linked and carried by the orbiting electrons. When the magnetic vortex systems, carriers of $q$ charge and $f$ flux quanta polarized along the magnetic field direction in the FQHE, cross a line parallel to the electric field there is a fixed relation between the charge and the linked flux crossing that line,

$$\frac{\Delta \Phi}{\Delta Q} = \frac{f\, h/2e}{qe} \quad . \tag{23}$$

If there are no resistive losses, the voltage induced along the line, by the carried induced flux cutting it, leads to a transverse resistance which is fractionally quantized,

$$R_t = \left( f/q \right) \left( h/2e^2 \right) \quad . \tag{24}$$

This is a fundamental general relation which only depends on the quanta $q, f$ carried by the carriers. This resistance value is a high precision number depending on fundamental constants and integers, which is displayed when appropriate microscopic conditions are met. The FQHE is such an experiment, which essentially measures the *ratio* of charge quanta to flux quanta of the carriers.

The generality of the argument may be turned around: The extraordinary accuracy of the measured fractional values indicates that nature is displaying new principles in these experiments and we may expect that current calculations techniques may require corrections. If this is the case, relations like the ones presented here may guide in improving the theory for the FQHE.

For the moment, we may claim that the experimental results are essentially compatible with the idea that the currents and voltages involved are produced by carriers of quanta of charge $e$ and magnetic flux $h/2e$ as indicated in the previous article.

1R: Laughlin, Phys. Rev. B23, 5652 (1981).

2J. K Jain, Phys. Rev. B41, 7653 (1990).

3Kamilla, Wu, J. K. Jain, Phys. Rev. Let., 76, 1332 (1996).

4G. González-Martín, Gen Rel. Grav. 23, 827 (1991); G. González-Martín, Physical Geometry, (Universidad Simón Bolívar, Caracas) (2000), Electronic copy posted at http:\\prof.usb.ve\ggonzalm\invstg\book.htm

5L. D. Landau, E. M. Lifshitz, Mécanique Quantique, Théorie non Relativiste (Ed. Mir, Moscow), 2nd. Ed. p. 496

(1965).

6J. D. Jackson, Classical Electrodynamics (John Wiley and Sons, New York), Second Ed., p. 589 (1975).

7D. Tsui, H. Stormer, A. Gossard, Phys. Rev. Lett. 48, 1559 (1982).and R. Willet, J. Eisenstein , H. Stormer , D. Tsui, A Gossard, J. English, Phys. Rev. Lett. 59, 1776 (1987).

8K. V. Klitzing, G. Dorda, M Pepper, Phys. Rev. Lett. 45, 494 (1980)

9R. Willet, R. Ruel, M. Paalanen, K. West, L. Pfeiffer, Phys. Rev. B 47, 7344 (1993).

10R. Du, H. Stormer, D. Tsui, L. Pfeiffer, K. West, Phys. Rev. Lett. 70, 2994 (1993).

11J. Eisenstein, L. Pfeiffer, K. West, Phys. Rev. Lett. 69, 3804 (1992).